\documentclass[a4paper]{jpconf}
\usepackage{graphicx}
\usepackage{epstopdf}
\usepackage{multicol}
\begin{document}
\title{Numerical methods for solution of the stochastic differential equations equivalent to the non-stationary Parker's transport equation
}

\author{A. Wawrzynczak$^{1}$, R. Modzelewska$^{2}$, M. Kluczek$^{2}$}

\address{$^{1}$Institute of Computer Sciences, Siedlce University, Poland, \\ $^{2}$Institute of Mathematics and Physics, Siedlce University, Poland.}

\ead{awawrzynczak@uph.edu.pl, renatam@uph.edu.pl}

\begin{abstract}
We derive the numerical schemes for the strong order integration of the set of the stochastic differential equations (SDEs) corresponding to the non-stationary Parker transport equation (PTE). PTE is 5-dimensional (3 spatial coordinates, particles energy and time)  Fokker-Planck type equation describing the non-stationary the galactic cosmic ray (GCR) particles transport in the heliosphere. We present the formulas for the numerical solution of the obtained set of SDEs driven by a Wiener process in the case of the full three-dimensional diffusion tensor. We introduce the solution applying the strong order Euler-Maruyama, Milstein and stochastic Runge-Kutta methods.  We discuss the advantages and disadvantages of the presented numerical methods in the context of increasing the accuracy of the solution of the PTE.
\end{abstract}

\section{Introduction}
\indent We employ the stochastic methodology to model the galactic cosmic rays (GCR) transport in the heliosphere.  Modulation of the GCR is a result of an action of four primary processes: convection by the solar wind, diffusion on irregularities of HMF, particles drifts in the non-uniform magnetic field  and adiabatic cooling. Transport of the GCR particles in the heliosphere can be described by the Parker transport equation  \cite{Pa}:
\begin{equation} \label{ParkerEq}
\small
\frac{\partial f}{\partial t}=\vec{\nabla}\cdot (K_{ij} ^{S}\cdot \vec{\nabla}f)-(\vec{v}_{d}+\vec{U})\cdot \vec{\nabla} f+\frac{R}{3}(\vec{\nabla} \cdot \vec{U})\frac{\partial f}{\partial R},
\end{equation}
\normalsize
where $f=f(\vec{r}, R, t) $ is an omnidirectional distribution function of three spatial coordinates $\vec{r}=(r,\theta,\varphi)$, particles rigidity $R$ and time $t$;  $\vec{U}$  is the solar wind velocity, $\vec{v}_{d}$  the drift velocity, and $K_{ij} ^{S}$ is the symmetric part of the diffusion tensor of the GCR particles.\\
\indent This paper is an extension of our previous results presented in \cite{WMG2015}. We have presented \cite{WMG2015} that the GCR transport can be effectively modelled based on the solution of the set of stochastic differential equations (SDEs) corresponding to the PTE (\ref{ParkerEq}). Firstly, the PTE must be brought to the form of the backward Fokker-Planck equation (e.g.\cite{Ga}), and then the corresponding SDEs must be solved numerically. In  \cite{WMG2015,Kopp2012} the Euler-Maruyama method was applied. In this paper, we increase the accuracy of the SDEs solution by applying the higher order methods i.e. Milstein and stochastic Runge-Kutta.

\section{Numerical methods to solve the stochastic differential equations}
The Euler-Maruyama method of solution  of the SDEs is unconditionally stable in the higher dimensions and doesn't rely on the numerical grid.
All integration method involves the statistical error, which can be reduced by increasing the number of simulated pseudoparticles and applying the higher order integration method e.g. Milstein's method or Runge-Kutta method e.g.\cite{RQ2014}. These three methods improve the approximation of the solution by applying higher order extension of the SDE solution in the Ito - Taylor series e.g.\cite{Kloeden}. The Euler-Maruyama method has order of convergence $\gamma=0.5$, addition just one more term from the Ito - Taylor expansion the order of convergence increases for the Milstein method up to $\gamma=1$. In turn, the stochastic Runge$-$Kutta  method requires to generate a new random variable $Z(t)$ resulting in strong order of convergence with $\gamma=1.5$. \\
\indent Let's consider a basic SDE:
\begin{equation}\label{Basic SDE}\small
dX_{t}=f(X_{t})dt+g(X_{t})dW_{t},
\end{equation}\normalsize
depending on the applied method the numerical approximation of the solution can be obtained by following formulas :
\small {\begin{eqnarray} \label{Ito - Taylor}
X_{j+1}&=&\underbrace{\underbrace{\underbrace{X_{j}+f\cdot dt+g\cdot dW_{t}}_{Euler - Maruyama}+\frac{1}{2}g\cdot g^{\prime}(dW_{j}^{2}-dt)}_{Milstein}+\Phi}_{Stochastic Runge - Kutta}
\end{eqnarray}}
where $\Phi=f^{\prime}\cdot g\cdot dZ_{j}+\frac{1}{2}(f\cdot f^{\prime}+\frac{1}{2}g^{2}\cdot f^{\prime\prime})dt^{2}+(f\cdot g^{\prime}+\frac{1}{2}g^{2}\cdot g^{\prime\prime})\cdot(dW_{j}\cdot dt-dZ_{j})+\frac{1}{2}g(g\cdot g^{\prime\prime}+g^{\prime 2})(\frac{1}{3}dW_{j}^{2}-dt)dW_{j}$
and $dZ_{i}=\frac{1}{2}dt(dW_{i}+\frac{dV_{i}}{\sqrt{3}})$ and $dV_{i}$ is an additional Wiener process e.g.\cite{Kloeden} .\\

\section{Stochastic differential equation corresponding to the Parker transport equation}
The PTE (Eq.\ref{ParkerEq}) in the 3-D heliocentric spherical coordinate system $(r,\theta,\varphi)$  written as time-backward FPE diffusion equation has the form:
\small
\begin{eqnarray}
\label{backwardParker}
\frac{\partial f}{\partial t}=A_{1}\frac{\partial^{2} f}{\partial r^{2}}+A_{2}\frac{\partial^{2} f}{\partial \theta^{2}}+A_{3}\frac{\partial^{2} f}{\partial \varphi^{2}}+A_{4}\frac{\partial^{2} f}{\partial r \partial \theta}+A_{5}\frac{\partial^{2} f}{\partial r \partial \varphi}+A_{6}\frac{\partial^{2} f}{\partial \theta \partial \varphi}+A_{7}\frac{\partial f}{\partial r}+A_{8}\frac{\partial f}{\partial \theta}+A_{9}\frac{\partial f}{\partial \varphi}+A_{10}\frac{\partial f}{\partial R}
\end{eqnarray}
\normalsize
the coefficients $A_{1}$,...,$A_{10}$ are presented in detail in \cite{WMG2015}. Depending on the choice  of the numerical approximation the corresponding to Eq.\ref{backwardParker} set of SDEs has a form:
\small
\begin{eqnarray}\label{SDE_all}
  dr &=& \underbrace{\underbrace{\underbrace{A_{7}\cdot dt+B_{11}\cdot dW_{r}}_{Euler - Maruyama}+\frac{1}{2}B_{11}\frac{\partial B_{11}}{\partial r}(dW_{r}^{2}-dt)}_{Milstein}+\Phi_{1}}_{Stochastic Runge - Kutta} \nonumber   \\
  d\theta &=& \underbrace{\underbrace{\underbrace{A_{8}\cdot dt+B_{21}\cdot dW_{r}+B_{22}\cdot dW_{\theta}}_{Euler-Maruyama}+\frac{1}{2}B_{21}\frac{\partial B_{21}}{\partial \theta}(dW_{r}^{2}-dt)+\frac{1}{2}B_{22}\frac{\partial B_{22}}{\partial \theta}(dW_{\theta}^{2}-dt)}_{Milstein}+\Phi_{2}}_{Stochastic Runge-Kutta}\\
 d\varphi &=& \underbrace{\underbrace{\underbrace{A_{9}\cdot dt+B_{31}\cdot dW_{r}+B_{32}\cdot dW_{\theta}+B_{33}\cdot dW_{\varphi}}_{Euler-Maruyama}+\Phi_{3}}_{Milstein}+\Phi_{4}}_{Stochastic Runge-Kutta}\nonumber \\
 dR &=& A_{10}\cdot dt. \nonumber
\end{eqnarray}
\normalsize
\begin{figure}[t]
  \begin{center}
\includegraphics[width=0.9\hsize]{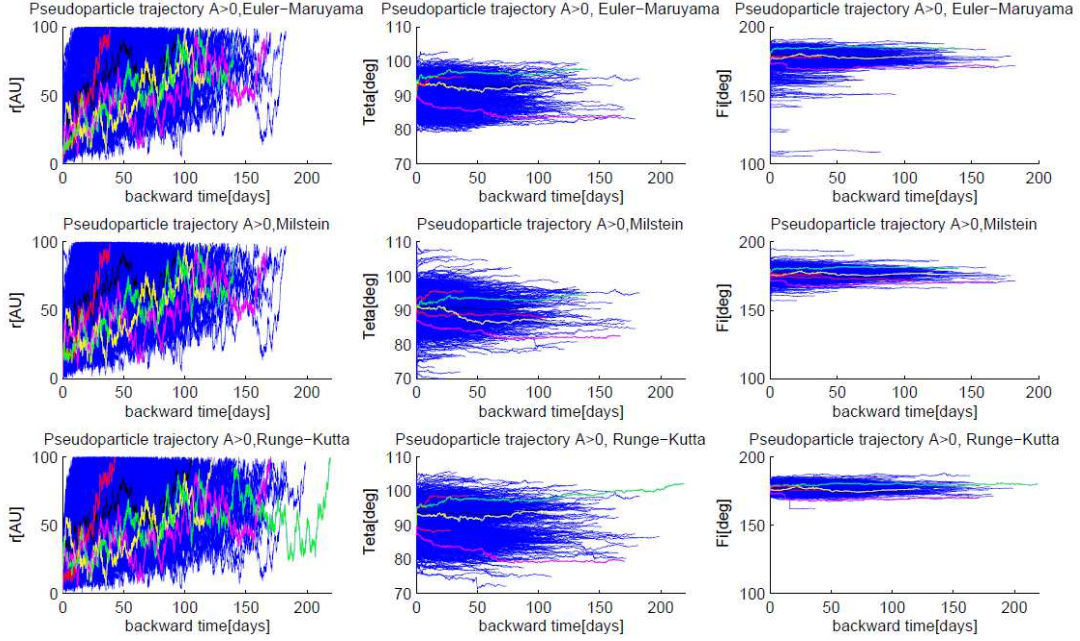}
\end{center}
\caption{\small{ \label{fig:fig1}Trajectories of the pseudoparticles with rigidity 10 GV for the A$>$0 solar magnetic cycle obtained applying the Euler-Maruyama, Milstein and Stochastic Runge-Kutta method. The specific colors highlight the trajectories of the sample pseudoparticles, based on the same Wiener process, traced backward in time from the heliosphere boundary until they reach the position $r=1AU$, $\theta=90^{\circ}$, $\varphi=180^{\circ}$.}}
\end{figure}

The $B_{ij}$, $(i,j=r,\theta,\varphi)$, is the lower triangular matrix presented in \cite{WMG2015}, and the coefficients $\Phi_{1}$, $\Phi_{2}$, $\Phi_{3}$ and $\Phi_{4}$ have the form:\\
\indent  $\Phi_{1}=B_{11}\cdot dZ_{r}\frac{\partial A_{7}}{\partial r}+\frac{1}{2}(A_{7}\frac{\partial A_{7}}{\partial r}+\frac{1}{2}B_{11}^{2}\frac{\partial^{2} A_{7}}{\partial r^{2}})dt^{2}+(A_{7}\frac{\partial B_{11}}{\partial r}+\frac{1}{2}B_{11}^{2}\frac{\partial^{2} B_{11}}{\partial r^{2}})(dW_{r}\cdot dt-dZ_{r})+\frac{1}{2}B_{11}(B_{11}\frac{\partial^{2} B_{11}}{\partial r^{2}}+(\frac{\partial B_{11}}{\partial r})^{2})(\frac{1}{3}dW_{r}^{2}-dt)dW_{r}$;

$\Phi_{2}=B_{21}\cdot dZ_{r}\frac{\partial A_{8}}{\partial \theta}+\frac{1}{2}(A_{8}\frac{\partial A_{8}}{\partial \theta}+\frac{1}{2}B_{21}^{2}\frac{\partial^{2} A_{8}}{\partial \theta^{2}})dt^{2}+(A_{8}\frac{\partial B_{21}}{\partial \theta}+\frac{1}{2}B_{21}^{2}\frac{\partial^{2} B_{21}}{\partial \theta^{2}})(dW_{r}\cdot dt-dZ_{r})+\frac{1}{2}B_{21}(B_{21}\frac{\partial^{2} B_{21}}{\partial \theta^{2}}+(\frac{\partial B_{21}}{\partial \theta})^{2})(\frac{1}{3}dW_{r}^{2}-dt)dW_{r}+B_{22}\cdot dZ_{\theta}\frac{\partial A_{8}}{\partial \theta}+\frac{1}{2}(A_{8}\frac{\partial A_{8}}{\partial \theta}+\frac{1}{2}B_{22}^{2}\frac{\partial^{2} A_{8}}{\partial \theta^{2}})dt^{2}+(A_{8}\frac{\partial B_{22}}{\partial \theta}+\frac{1}{2}B_{22}^{2}\frac{\partial^{2} B_{22}}{\partial\theta^{2}})(dW_{\theta}\cdot dt-dZ_{\theta}) +\frac{1}{2}B_{22}(B_{22}\frac{\partial^{2} B_{22}}{\partial \theta^{2}}+(\frac{\partial B_{22}}{\partial \theta})^{2})(\frac{1}{3}dW_{\theta}^{2}-dt)dW_{\theta} $;

$\Phi_{3}=\frac{1}{2}B_{31}\frac{\partial B_{31}}{\partial \varphi}(dW_{r}^{2}-dt)+\frac{1}{2}B_{32}\frac{\partial B_{32}}{\partial \varphi}(dW_{\theta}^{2}-dt)+\frac{1}{2}B_{33}\frac{\partial B_{33}}{\partial \varphi}(dW_{\varphi}^{2}-dt)$;

$\Phi_{4}=B_{31}\cdot dZ_{r}\frac{\partial A_{9}}{\partial \varphi}+\frac{1}{2}(A_{9}\frac{\partial A_{9}}{\partial \varphi}+\frac{1}{2}B_{31}^{2}\frac{\partial^{2} A_{9}}{\partial \varphi^{2}})dt^{2}+(A_{9}\frac{\partial B_{31}}{\partial \varphi}+\frac{1}{2}B_{31}^{2}\frac{\partial^{2} B_{31}}{\partial \varphi^{2}})(dW_{r}\cdot dt-dZ_{r})+\frac{1}{2}B_{31}(B_{31}\frac{\partial^{2} B_{31}}{\partial \varphi^{2}}+(\frac{\partial B_{31}}{\partial \varphi})^{2})(\frac{1}{3}dW_{r}^{2}-dt)dW_{r}+B_{32}\cdot dZ_{\theta}\frac{\partial A_{9}}{\partial \varphi}+\frac{1}{2}(A_{9}\frac{\partial A_{9}}{\partial \varphi}+\frac{1}{2}B_{32}^{2}\frac{\partial^{2} A_{9}}{\partial \varphi^{2}})dt^{2}+(A_{9}\frac{\partial B_{32}}{\partial \varphi}+\frac{1}{2}B_{32}^{2}\frac{\partial^{2} B_{32}}{\partial \varphi^{2}})(dW_{\theta}\cdot dt-dZ_{\theta})+\frac{1}{2}B_{32}(B_{32}\frac{\partial^{2} B_{32}}{\partial \varphi^{2}}+(\frac{\partial B_{32}}{\partial \varphi})^{2})(\frac{1}{3}dW_{\theta}^{2}-dt)dW_{\theta}+B_{33}\cdot dZ_{\varphi}\frac{\partial A_{9}}{\partial \varphi}+\frac{1}{2}(A_{9}\frac{\partial A_{9}}{\partial \varphi}+\frac{1}{2}B_{33}^{2}\frac{\partial^{2} A_{9}}{\partial \varphi^{2}})dt^{2}+(A_{9}\frac{\partial B_{33}}{\partial \varphi}+\frac{1}{2}B_{33}^{2}\frac{\partial^{2} B_{33}}{\partial \varphi^{2}})(dW_{\varphi}\cdot dt-dZ_{\varphi})+ \frac{1}{2}B_{33}(B_{33}\frac{\partial^{2} B_{33}}{\partial \varphi^{2}}+(\frac{\partial B_{33}}{\partial \varphi})^{2})(\frac{1}{3}dW_{\varphi}^{2}-dt)dW_{\varphi} $.
\section{Results and Summary}
We performed the simulation applying all three methods given by Eqs.\ref{SDE_all}. The trajectory of $n=3000$ pseudoparticles was traced backward in time in the spherical heliocentric coordinate system. The pseudoparticles were initialized in the point representing the Earth's orbit (i.e. $r=1AU$, $\theta=90^{\circ}$, $\varphi=180^{\circ}$ ) and  traced backward in time until crossing the heliosphere boundary assumed at 100 AU (see Fig. 1 in \cite{WMG2015}). The value of the particle distribution function $f(\vec{r}, R)$ for the starting point was be found as an average of $f_{LIS}(R)$ value for  pseudoparticles characteristics at the entry positions, $f(\vec{r}, R)=\frac{1}{N}\sum_{n=1}^{N}f_{LIS}(R)$,
where $f_{LIS}(R)$ is the cosmic ray local interstellar spectrum taken as in \cite{Webber} for rigidity $R$ of the $n^{th}$ particle at the entrance point.\\
\begin{figure}[t]
  \begin{center}
\includegraphics[width=0.75\hsize]{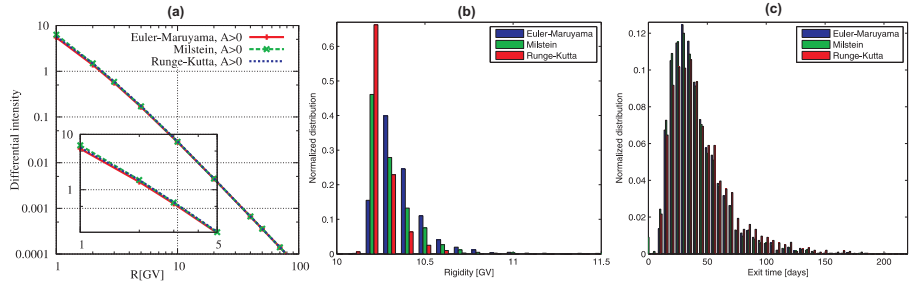}
\end{center}
\caption{\small{\label{fig:figall} (a)Simulated galactic protons rigidity spectra and the histograms of (b) the particles rigidity and (c) the particles exit time for the pseudoparticles initialized R=10 GV from position $r=1AU$, $\theta=90^{\circ}$, $\varphi=180^{\circ}$ for the A$>$0 solar magnetic cycle obtained applying the Euler-Maruyama, Milstein and Stochastic Runge-Kutta method.}}
\end{figure}
\indent In the case when we know the analytical solution it is easy to compare the efficiency of each numerical method. However, here we do not have such a possibility. Thus, we have compared the simulated galactic protons rigidity spectra for  A$>$0 solar magnetic cycle. Fig.~\ref{fig:figall}a presents that all applied numerical schemes resulted in the same spectra, slight differences are seen for the lower rigidity particles. The obtained spectrum is in agrement with results presented in \cite{Zhang1999}.\\
\indent Fig.~\ref{fig:fig1} presents the trajectories of the pseudoparticles with rigidity 10 GV for the A$>$0 solar magnetic cycle with respect to all coordinates. To allow the reliable comparison all simulations were based on the same Wiener process. One can see that there arises some subtle differences: the trajectories of pseudoparticles vs. the heliolongitude are the widest for Euler-Maruyama and narrow for the Runge-Kutta method.
At the same time the opposite distribution  is observed   vs. the heliolatitude. Additionally, we have estimated the histograms of the  particles rigidity and the particles exit time for all three methods. Fig.~\ref{fig:figall}bc show that the time of traveling the pseudoparticles throughout the heliosphere is the longest applying the stochastic Runge-Kutta, at the same time particles loses less energy during travel. \\
\indent The obtained results suggest that all applied methods give a reliable results. Performed tests proved that the Milstein and stochastic Runge-Kutta methods are more stable and return the same values of differential spectra when we decrease the number of simulated particles by factor of three. However, the slight differences between the pseudoparticles trajectories and its characteristics (rigidity and exit time) given by the tested numerical schemes must be analyzed in more detail.

\section{Acknowledgments and References}
\small {This work is supported by The Polish National Science Centre grant awarded by decision number DEC-2012/07/D/ST6/02488.}

\end{document}